\documentclass[aps,pre,superscriptaddress,twocolumn]{revtex4}

\usepackage{graphicx}
\usepackage{amssymb,latexsym,amsthm,amsmath}
\usepackage{float}
\begin{document}

\title{Synchronization interfaces and overlapping communities in complex networks}
\author{D. Li}
\affiliation{Department of Physics, Minerva Center, Bar Ilan
University, Ramat Gan 52900, Israel}
\author{I. Leyva}
\author{J.A. Almendral}
\author{I. Sendi\~na--Nadal}
\author{J.M. Buld\'u}
\affiliation{Dpto. de F\'{\i}sica, Universidad Rey Juan Carlos, c/ Tulip\'an s/n,
28933 M\'ostoles, Madrid, Spain}
\author{S. Havlin}
\affiliation{Department of Physics, Minerva Center, Bar Ilan
University, Ramat Gan 52900, Israel}
\author{S. Boccaletti}
\affiliation{Embassy of Italy in Tel Aviv, 25 Hamered St., 68125 Tel
Aviv, Israel}
\affiliation{CNR- Istituto dei Sistemi Complessi, Via Madonna del Piano, 10, 50019 Sesto Fiorentino (Fi), Italy}

\date{\today}

\begin{abstract}
We  show that a complex network of phase oscillators may display
interfaces between domains (clusters) of synchronized oscillations.
The emergence and dynamics of these interfaces are studied in the
general framework of interacting phase oscillators composed of
either dynamical domains (influenced by different forcing
processes), or structural domains (modular networks). The obtained
results allow to give a functional definition of overlapping
structures in modular networks, and suggest a practical method to
identify them. As a result, our algorithm could detect information
on both single overlapping nodes and overlapping clusters.

\end{abstract}

\maketitle

The functioning of many natural (biological, neural, chemical) or
artificial (technological) networks displays coordination of
parallel tasks \cite{review}. This phenomenon may be represented as
the interplay between two simultaneous processes. The first
(involving most of the network nodes) leads to the emergence of
organized clusters (or moduli, or cohesive subgroups), where nodes
in the same cluster adjust their dynamics into a common
(synchronized) behavior to enhance the performance of a specific
task. The second process (involving just few nodes of the graph) is
to form interfaces (or overlapping structures) between the moduli,
that are responsible for the coordination between the different
tasks.

A practical example is the brain functioning, wherein the vision of
the images from the right and left eye can be represented as the
emergence of a collective behavior in different, close-by areas of
neurons, whereas perception (as e.g. feature binding) implies the
coordination of the two emergent dynamics \cite{varela}. Another
example in social science is, for instance, consensus formation in
polarized elections (as e.g. between a left-wing and a right-wing
candidate for the presidency), where usually most of citizens cast
their votes following a consistent opinion (they always vote for the
left-wing or the right-wing party), whereas usually a small fraction
(mostly being responsible for the final election's outcome)
alternate in time their preference depending on their actual opinion
\cite{social}.

In this Letter, we present the first evidence that, under the
presence of different  functional  (synchronized) clusters,
interfaces appear and show a novel specific dynamical behavior. This
finding enables us to develop an algorithm for extracting
information on the overlapping structure in a modular network.
Indeed, the study of separate modular structures \cite{modular} and
synchronization \cite{synchro} in complex graphs has not so far
unraveled the crucial point concerning the role of synchronization
interfaces and its usefulness in detecting overlapping communities.

In the following, without lack of generality, we study networks
consisting of two domains of interacting phase oscillators (each one
synchronously evolving at a different frequency), where the nature
of the two different frequency domains is the result of either  a
dynamical process (influenced by different forcing processes) or a
structural design (modular network). Under these conditions, at each
time, most of the oscillators will contribute to the synchronous
behavior of the two clusters, whereas a few nodes will find
themselves in a {\it frustrating} situation of having to decide how
to behave as a consequence of the contrasting inputs received by the
two clusters.

Let us first describe the case of a synchronization interface
emerging from a dynamical process in a generic random graph $G$ of
$N$ coupled oscillators, whose original frequencies $\{
\omega_{i}\}$ are randomly drawn from a uniform distribution in the
interval $0.5\pm 0.25$, subjected simultaneously to an internal
bidirectional coupling and an external pace-making unidirectional
forcing. The network dynamics is described by:

\begin{eqnarray}
\dot\phi_i=\left\{
 \begin{array}{llll}
\omega_{i} & + & \displaystyle
\frac{d}{(k_i+k_{p_i})}\displaystyle \sum_{j=1}^{N}
a_{ij}\sin(\phi_j-\phi_i) &  \\
& + & \displaystyle \frac{d_p k_{p_i}}{(k_i+k_{p_i})}\displaystyle
\sin(\phi_{p_i}-\phi_i), & \label{eq:system}
  \end{array}\right.
\end{eqnarray}
\noindent where dots denote temporal derivatives, $k_i$ is the
degree of the $i^{th}$ oscillator, $\phi_{p_i}$ is  the
instantaneous phase of a forcing oscillator having  $k_{p_i}$
unidirectional connections, $d$ and $d_p$ are coupling strengths,
and $\{a_{ij} \}$ are either $1$ or $0$ depending on whether or not
a link exists between node $i$ and node $j$.

In our simulation, we study a network $G$ which consists of $N=200$
phase oscillators arranged in an Erd\"os-R\'enyi configuration
\cite{erdos}. Initially, we set $d_p=0$ and $k_{p_i}=0 \ \forall i$,
and we choose $d=0.1$ such that $G$ features an unsynchronized
motion. Next, we arbitrarily divide the nodes into two groups: nodes
from $i=1,\cdots,100$ (from $i=101,\cdots,200$) are assigned to the
community $A$ (to the community $B$). We introduce two pacemakers of
frequencies $\omega_{p_A}$ and $\omega_{p_B}$, and connect the nodes
in the first group (in the second group) with the first (second)
pacemaker. This implies in Eq.~(\ref{eq:system}) to set $\phi_{p_i}=
\phi_{p_A} \equiv \omega_{p_A} t $ ($\phi_{p_i}= \phi_{p_B} \equiv
\omega_{p_B} t $) for all the nodes in $A$ ($B$). In order to assign
to each node its $k_{p_i}$ links with the pacemaker, we start at
$t_0=0$ the evolution of Eq.~(\ref{eq:system}) from random initial
conditions in the unforced case ($k_{p_i}=0 \ \forall i $), and add
at later times $t_l=t_0+l\Delta t$ links between nodes of the two
communities and the pacemakers. Precisely, at each time $t_l$, the
pace-maker $\phi_{p_A}$ ($\phi_{p_B}$) forms a connection with that
node $j$ in $A$ ($B$) whose instantaneous phase at time $t_l$
corresponds to the minimum  $ {\min}_j \left | \delta -
\Delta\theta_j \bmod
 2\pi \right |, $ with
  $\Delta\theta_j=\phi_j(t_l)-\phi_{p}(t_l)$, $\delta \in (0, 2
  \pi)$ a suitable parameter, and $\phi_{p}(t_l)=\phi_{p_A}(t_l)$
  ($\phi_{p}(t_l)=\phi_{p_B}(t_l)$) for those nodes in $A$ ($B$).
Ref. \cite{isendinapre08} demonstrated that, by operating this
attachment over a given time interval, the resulting dynamics of
any arbitrary network of oscillators can be entrained to any
arbitrary frequency $\omega_{p}$.

\begin{figure}
 \centering
\includegraphics[width=0.3\textwidth]{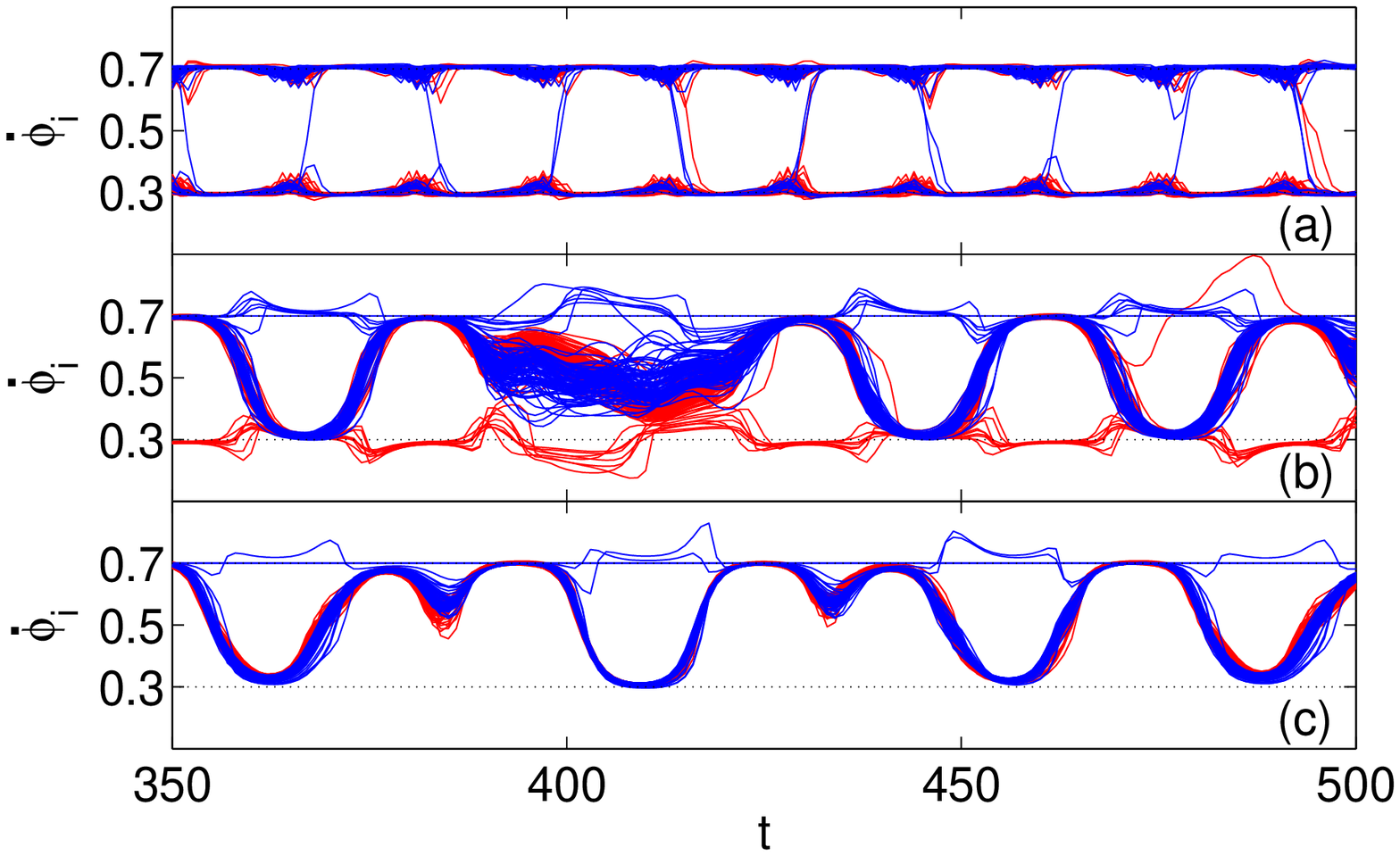}
\includegraphics[width=0.25\textwidth]{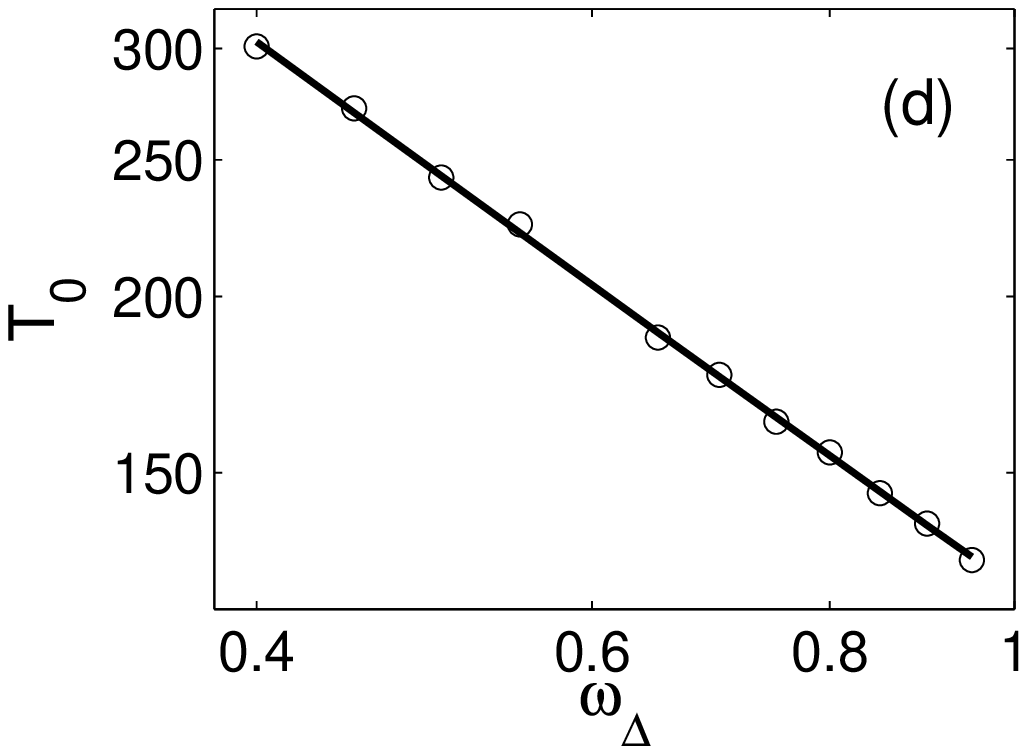}
\caption{(Color online) (a-c) Instantaneous frequencies $\dot \phi_i (t)$ of each
one of the 100 oscillators in $A$ (red lines)  [in $B$, (blue
lines)] vs. time, obtained from simulation of
Eq.~(\ref{eq:system}) for (a) $d = 3.25$, (b) $d = 5.75$, and (c)
$d = 9.75$. (d) Log-log plot of the period $T_O$ of the switching
process in the interface vs. $\omega_{\Delta}$, for $d = 3.5$. The
solid line represents a linear fit with $T_O \sim 120 /\omega_{\Delta}$.
 Each point is the average of 5 independent realizations.
Other parameters reported in the text. \label{fig1}}
\end{figure}

By selecting $\omega_{p_A}=0.7$, $\omega_{p_B}=0.3$, $\sum_{i=1}^N
k_{p_i}=2,000$ links to the pacemakers (1,000 to entrain the nodes
in $A$ and the other 1,000 to entrain those in $B$), and $d_p=1$,
the dynamics displays two large communities of entrained
oscillators, densely intermingled by the connectivity of $G$. The
situation is depicted in Fig.~\ref{fig1}(a) where we report the
instantaneous frequency of each oscillator in $G$ (averaged over a
small window to smooth fluctuations) as a function of time. We
observe that most of the nodes in $A$ (in $B$) have a constant
frequency (that of the corresponding
 pacemaker), whereas the few nodes belonging to the
synchronization interface  feature a frequency which is
oscillating around the mean value of that of the two communities
$\bar \omega= (\omega_{p_A}+\omega_{p_B})/2$. Furthermore,
Fig.~\ref{fig1}(d) gives evidence that the period $T_O$ of the
switching process in the interface is inversely proportional to
$\omega_{\Delta}=(\omega_{p_A}-\omega_{p_B})/2$.

Notice that this switching mechanism is the result of the
competition of two conflicting processes: the synchronization within
$G$ (controlled by the parameter $d$) that would lead the whole
network to exhibit a unique frequency, and the forcing of the two
pacemakers (controlled by $d_p$) that would try to separate the
nodes into two clusters of entrained oscillators. In order to
quantify this competition and to describe the size of the
synchronization interface we fix $d_p=1$ and gradually increase $d$.
The results are shown in Figs.~\ref{fig1}(a-c) with (a) $d=3.25$,
(b) $d=5.75$,  and (c) $d=9.75$. For intermediate coupling
Fig.~\ref{fig1}(b), it is observed that the interface grows in size,
recruiting more and more nodes out of the two communities into the
oscillating mode. Due to the presence of the two forcing pacemakers,
this interface is organized in an oscillating mode rather than in a
constant frequency mode, as it would occur in the classical Kuramoto
regime \cite{kura}. As the coupling $d$ is further increased, almost
the entire system of oscillators eventually participates into this
interface oscillating mode, as seen in Fig.~\ref{fig1}(c). Notice
that similar switching dynamics was observed before in the case of a
chain of oscillators subjected to two forcing frequencies applied to
the two ends of the chain \cite{inma}.

In order to give an analytical insight to our findings, let us
consider the simple case of three interacting phase oscillators
described by: $ \dot \phi_1=\omega_1+K_1 \sin (\phi_3-\phi_1) \ ;
\dot \phi_2=\omega_2+K_2 \sin (\phi_3-\phi_2) \ ; \dot
\phi_3=\omega_3+K [ \sin (\phi_1-\phi_3) + \sin (\phi_2-\phi_3) ]$.
Here,  $\phi_3$ is the phase of an oscillator (with natural
frequency $\omega_3$) receiving a simultaneous coupling from two
other oscillators at natural frequencies $\omega_1 \neq \omega_2
\neq \omega_3$, and $K_1,K_2 \ll K$ are coupling constants. The
oscillators 1 and 2 model the two functional (frequency) domains $A$
and $B$, where the nature of the two different frequency domains is
from either a dynamical process or a structural design. These two
domains of synchronous oscillators simultaneously interact with the
small group of nodes in the interface (modeled by the third
oscillator), so that we can reasonably assume $K_1=K_2=0$, and
consequently $\phi_{1,2}=\omega_{1,2} t$.

The resulting equation for $\phi_3$

\begin{equation}
\dot \phi_3= \omega_3 +2K \sin \left( \frac{\omega_1+\omega_2}{2}
t -\phi_3 \right) \cos \left( \frac{\omega_1-\omega_2}{2} t
\right)
  \label{equ1}
\end{equation}
\noindent yields an analytic solution

\begin{equation}
\dot \theta_3= \frac{\tilde A e^{\frac{2K}{\omega_\Delta} \sin
(\omega_\Delta t)} \cos(\omega_\Delta t)}{1+\tilde B
e^{\frac{4K}{\omega_\Delta}\sin (\omega_\Delta t)}} \ ,
  \label{equ2}
\end{equation}
\noindent for $\omega_3=\bar \omega$, and where
$\theta_3=\phi_3-\bar \omega t$, $\bar \omega=
(\omega_1+\omega_2)/2$, $\omega_\Delta= (\omega_1-\omega_2)/2$, and
$\tilde A, \tilde B$ suitable parameters. In good agreement with the
results of Fig.~\ref{fig1}(d),  Eq.~(\ref{equ2}) predicts that the
instantaneous frequency $\dot \phi_3$ (once $\omega_3$ is selected
to be the mean frequency of the two forcing clusters) will oscillate
around $\bar \omega$ with a period $T_O$ {\it inversely
proportional} to the difference in the frequencies of the two
forcing clusters. Notice that, as $K$ increases (i.e. in the strong
coupling regime), $\dot \theta_3$ will progressively vanish, which
means that $\dot \phi_3 = \bar \omega$ or, in other words, the
frequency of the oscillator will be locked to the mean frequency of
the two giant clusters.

While so far we have considered the behavior of interfaces as the
result of the competition of dynamical domains, we now move to
describe the case of synchronization interfaces under the
competition of structural domains in modular graphs in the absence
of the forcing [$d_p=0$ and $k_{p_i}=0 \ \forall i$ in
Eq.~(\ref{eq:system})]. For this purpose, we construct the adjacency
matrix of $G$ by considering two large communities ($A$, $B$), each
one formed by 50 densely and randomly connected nodes (the average
degree in the same community is 16), which are overlapped by a small
community $O$ made of a complete graph of 5 nodes($3$ links to $A$
($B$)) that form symmetric connections to nodes in both $A$ and $B$.
Each of the 105 nodes of $G$ is associated with a phase oscillator
obeying Eq.~(\ref{eq:system}), which is integrated with a
Runge-Kutta-Fehlberg  method, for an initial distribution of
frequencies such that nodes in $A$ ($B$) [i.e. nodes from $i=1$
($i=51$) to  $i=50$ ($i=100$)] have natural frequencies uniformly
distributed in  the range $0.25\pm 0.25$ ($0.5\pm0.25$), while nodes
in $O$ have frequencies uniformly distributed in the range
$0.375\pm0.05)$ (i.e. around the mean frequency of the two
distributions).

\begin{figure}
 \centering
\includegraphics[width=0.45\textwidth]{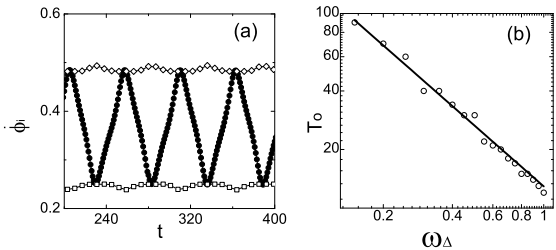}
\caption{(a) Instantaneous frequencies $\dot \phi_i (t)$  vs. time
from simulation of Eq.~(\ref{eq:system}) with $d=0.1$ (other
parameters and stipulations are reported in the text). Squares,
diamonds, and full circles represent respectively nodes belonging to
$A$, $B$ and $O$. (b) Log-log plot of the switching period $T_O$ of
the oscillations in the frequency of the nodes in $O$ vs. the
frequency difference $\omega_\Delta$. The solid line represents a
linear fit with slope $-1.002 \pm 0.0069$. \label{fig2}}
\end{figure}

Figure~\ref{fig2} shows that all oscillators in the communities $A$
and $B$ behave synchronously with an almost constant frequency in
time (well approximating the mean of the original frequency
distribution), while all oscillators in $O$ constitute the
synchronization interface and, as so, display an instantaneous
frequency oscillating in time around the mean value of the two
frequencies in the two clusters [Fig. \ref{fig2}(a)]. Similarly,
Fig. \ref{fig2}(b) shows the good agreement between the analytical
prediction of Eq.~(\ref{equ2}) and the numerical results concerning
the scaling of the period of the frequency oscillations of $O$ with
the frequency difference between the two communities $A$ and $B$.

\begin{figure}
 \centering
\includegraphics[width=0.5\textwidth]{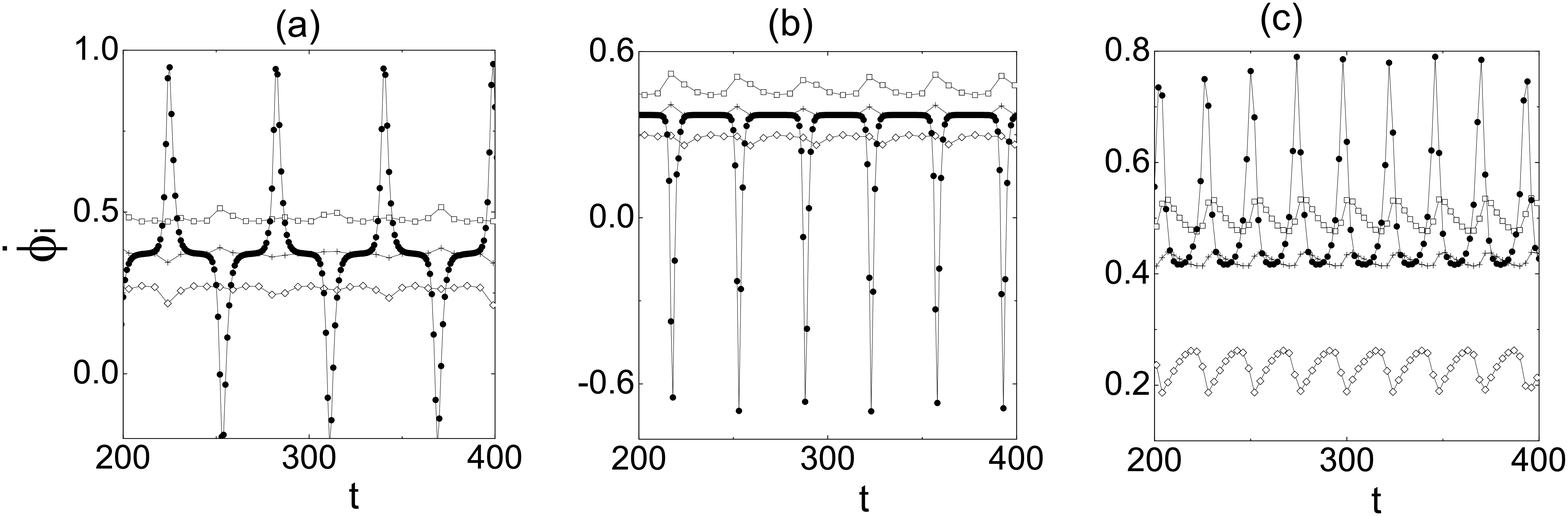}
\caption{Effects of coupling strength (a-b) and asymmetry (c). (a)
$d=0.5$; (b) $d=0.95$. (c) $d=0.1$, but 2 (5) links of the nodes
in the interface are to nodes of cluster $A$ ($B$). As in
Fig.~\ref{fig2},  squares, diamonds and full circles refer
respectively to nodes belonging to $A$, $B$ and $O$. \label{fig3}}
\end{figure}

It is important to remark that all other analytical predictions of
Eq.~(\ref{equ2}) concerning the behavior of the synchronization
interface in the large coupling regime can be confirmed in our
simulations. Indeed, looking at Fig.~\ref{fig3}(a-b) one realizes
that the effect of increasing the coupling strength $d$ is to lock
almost always the synchronized frequency of the interface to the
mean of the frequencies of the two communities, with persistent
events of short shootings to even larger (or lower) frequencies.

What described so far refers to the specific case in which initially
the nodes of the interface were prepared to have the same number of
links to nodes of the two communities. It is therefore interesting
to ask what happens when the nodes in $O$ are asymmetrically
connected to $A$ and $B$. While the low coupling regime does not
substantially differ from the symmetrical case, the high coupling
regime [illustrated in Fig.~\ref{fig3}(c)] features frequency
oscillations of the nodes in $O$ that are biased toward that
community to which the nodes have more connections. In fact, it can
be proved that if the nodes in $O$ have $k_i^A$ ($k_i^B$)
connections to nodes in $A$ ($B$), they tend to lock to the weighted
mean frequency $\bar\omega_w=\frac{k_i^A \omega_A + k_i^B
\omega_B}{k_i^A + k_i^B}$, where $\omega_A$ ($\omega_B$) is the
average frequency of the nodes in $A$ ($B$).

\begin{figure}
\includegraphics[width=0.23\textwidth]{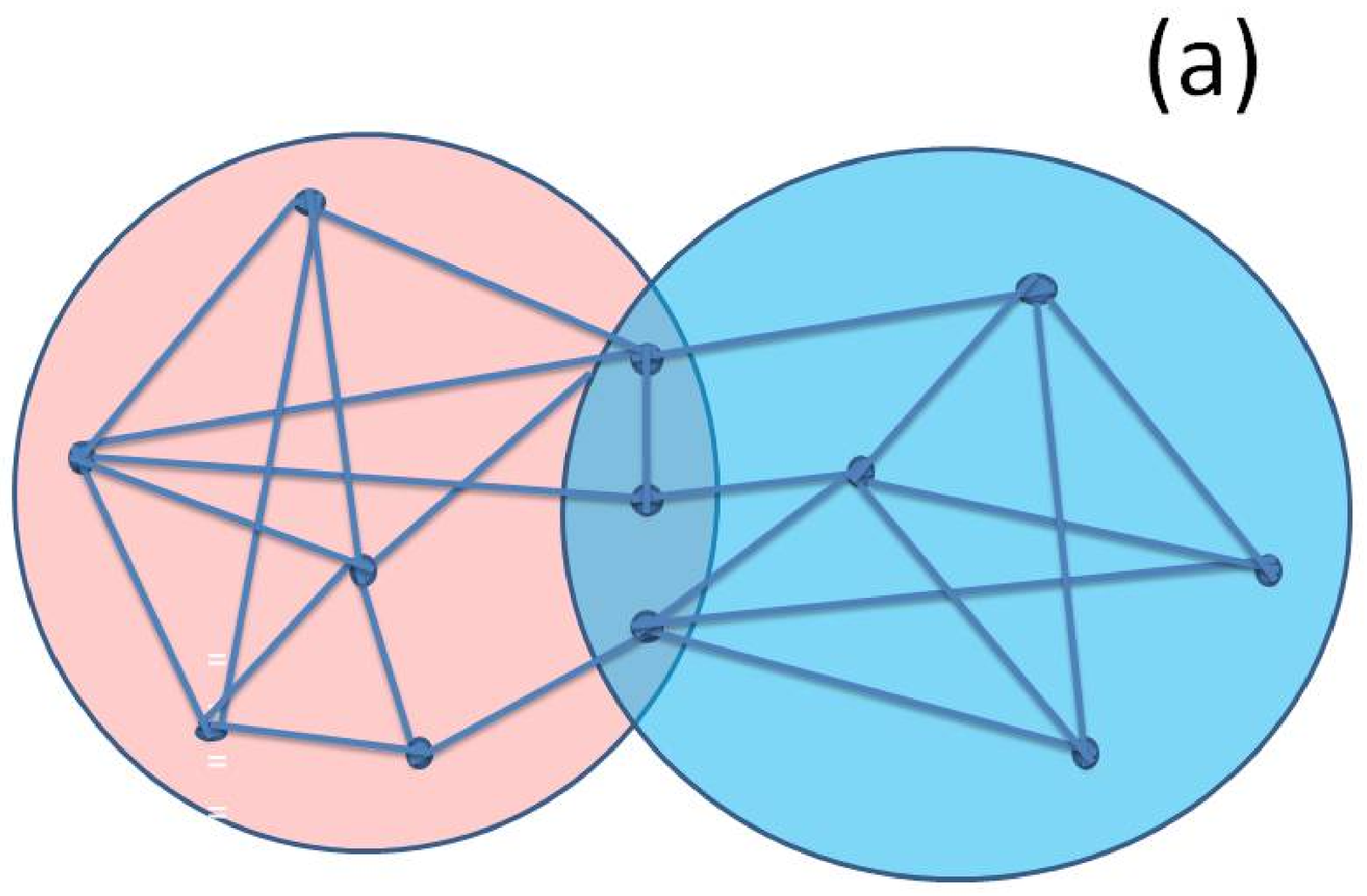}\includegraphics[width=0.25\textwidth]{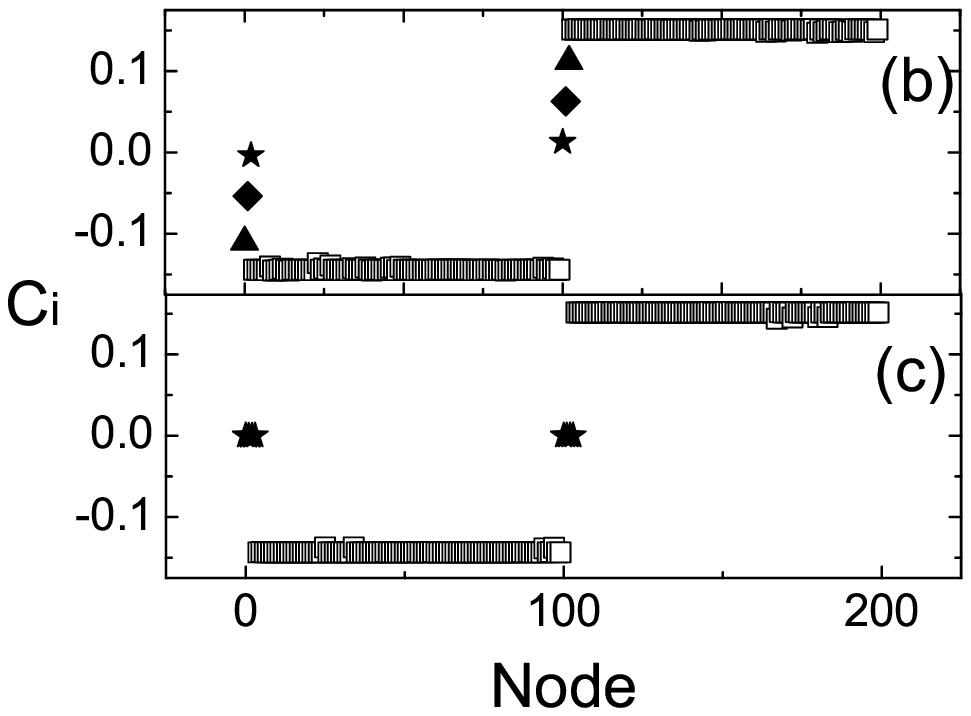}
\caption{(a) Illustration of the concept of overlapping structure,
where the overlapping region of two circles (communities) represents
the overlapping community.
 (b) $C$ (see text for definition) vs. node index in $G$. Overlapping
nodes $1,2,3$($101,102,103$), labeled respectively with triangle,
diamond and star, have $[1,5], [2,4], [3, 3]$ connections with nodes
of [$B$, $A$]([$A$, $B$]), while $4-100 (104-200)$ labeled with
squares belong functionally to $A$ ($B$) and $d=0.15$ (c)
Overlapping clusters each consisting of complete graphs of nodes
$1-4 (101-104)$, labeled with stars, have symmetrical connections
with $A$ and $B$, while nodes $5-100 (105-200)$, labeled with
squares, belong functionally to $A$ ($B$), here $d=0.1$.
\label{fig4}}
\end{figure}

The overall scenario reported above suggests a practical way to
detect overlapping communities (Fig.~\ref{fig4}(a)) in generic
modular networks. It is essential to remark that most of the
definitions of network communities proposed so far led essentially
to a graph partition into components, such that a given node belongs
to and only to one of the components of the partition
\cite{modular}. The possibility, instead, that two components of a
partition can have an overlapping set of nodes has been recently
investigated by means of topological arguments \cite{vicsek}. The
novelty of our approach is in introducing a functional concept of
overlapping structures, that are defined in relationship to
dynamical response of the network as a whole. Namely, as far as
synchronized behavior of phase oscillators is concerned, we define
an overlapping structure as the set of nodes which, instead of
following the constant frequency of one of the two domains, balance
their instantaneous frequencies in between these two, and as so,
they cannot be considered as a functional part of any single domain.
As we will see below, this definition allows detection of further
information including single overlapping nodes, as compared to
previous studies.

To illustrate this idea, we construct a network $G$ made of two
large moduli ($A$ and $B$ of 100 nodes each), where the majority of
nodes forms random connections (with average degree 15) with only
elements of the same community, while only very few nodes form links
with nodes of both communities. Precisely, we call $k_i^A$ ($k_i^B$)
the total number of links these few nodes form with nodes in $A$
($B$), and define a quantity to evaluate the degree of overlapping
of these nodes as the ratio $ D_i=k_i^{A}/k_i^{B}$. When the node is
perfectly overlapped between two clusters, $D_i=1$. Under these
conditions, Eq.~(\ref{eq:system}) is simulated for $d_p=0$,
$k_{p_i}=0\ \forall i$, the set $\{\omega_i\}$ drawn from a Gaussian
distribution with standard deviation $0.1$ and mean value $0.3$
($0.6$) for nodes belonging to community $A$ ($B$). To identify
those overlapping nodes, the quantity
$C_i=sgn[\dot\phi_i(t)-\bar{\omega}]\min_t\{|\dot\phi_i(t)-\bar{\omega}|\}$
($\bar\omega$ is the mean of the two averaged frequencies assigned
to the two communities) is introduced, which allows to monitor how
close in time the dynamics of a node gets to $\bar\omega$. When the
node is in the dynamics of fully synchronization interface, {\bf
$C_i= 0$}. Therefore, the closer $D_i$ gets to $1$, the closer $C_i$
is expected to get to $0$.  The results are shown in
Fig.~\ref{fig4}(b,c) for two different arrangements of the
overlapping community: overlapping nodes [Fig.~\ref{fig4}(b)] and
overlapping clusters [Fig.~\ref{fig4}(c)] which have symmetrical
connections to two clusters. In both cases, two large synchronized
clusters are identified very far from the overlapping
synchronization, corresponding to those nodes performing distinct
tasks and already classified by the structural partition. At the
same time, the dynamical evolution manifests a group of nodes whose
dynamics is located significantly out of the two main clusters (thus
identifying the overlapping community). In Fig.~\ref{fig4}(b), each
overlapping node gives rise to a different value of $C_i$ in
correspondence to its specific degree of overlapping $D_i$. Namely,
the node with $D_i=1$ yields $C_i=0$. On the contrary,  in
Fig.~\ref{fig4}(c), all the nodes inside the overlapping cluster are
identified as a whole and feature the same value of $C_i=0$ due to
symmetrical connection.

In conclusion, we have analytically and numerically described the
behavior of synchronization interfaces in both random and modular
complex networks of phase oscillators. We identify the overlapping
structure (nodes or clusters) in the modular network, using for the
first time a functional approach. Therefore, our study is of
practical relevance for applications to large biological, neural,
chemical, social and technological modular networks, where, besides
detection of the structure of overlapping communities, one is
interested in inspecting the coordinating role of such overlapping
communities in the functional performance of the graph.

Work partly supported by EU contract 043309 GABA, by the Spanish
Ministry of S\&T under Project n. FIS2006-08525, and by the URJC-CM
under Projects n. URJC-CM-2006-CET-0643 and 2007-CET-1601. We also
wish to thank the EU project DAPHNet, ONR, the Israel Science
Foundation, Hadar Foundation and the  Center of Complexity Science
for financial support.


\end{document}